\documentclass[twocolumn,10pt]{revtex4-1} 
\usepackage{graphicx,color,transparent}
\usepackage{amsmath,amssymb,braket,verbatim}

\begin{document}

\title{Electromagnetically induced absorption in a non-degenerate three-level ladder system}

\author{Daniel J Whiting}\email{daniel.whiting@durham.ac.uk}
\author{Erwan Bimbard}
\author{James Keaveney}
\author{Mark A Zentile}
\author{Charles S Adams}
\author{Ifan G Hughes}

\affiliation{Joint Quantum Centre (JQC) Durham-Newcastle, Department of Physics, Durham University, South Road, Durham, DH1 3LE, UK}

\begin{abstract}
We investigate, theoretically and experimentally, the transmission of light through a thermal vapour of three-level ladder-type atoms, in the presence of 2 counter-propagating control fields.
A simple theoretical model predicts the presence of electromagnetically induced absorption (EIA) in this pure three-level system when the control field is resonant.
Experimentally, we use $^{87}$Rb in a large magnetic field of 0.62~T to reach the hyperfine Paschen-Back regime and realise a non-degenerate three-level system.
Experimental observations verify the predictions over a wide range of detunings.
\end{abstract}


\maketitle

The study of coherent phenomena in driven multi-level atomic systems is an active area of research \cite{Fleischhauer2005}.
Three-level atoms driven by two applied fields display a variety of effects, including electromagnetically induced transparency (EIT) \cite{Harris1997} and coherent population trapping \cite{Gray1978}. By adding further fields and states, different phenomena have been observed such as the appearance of electromagnetically induced gratings/Bragg reflection \cite{Kuang2008,Babin2003,Bae2010} and electromagnetically induced absorption (EIA) \cite{Akulshin1998,Lezama1999a}.
In contrast to the sharp increase in resonant transmission that characterises EIT, these effects are identified by a resonant transmission that can be decreased by the presence of additional control fields.
A concomitant change in sign of dispersion can be used to switch between subluminal and superluminal light propagation \cite{Agarwal2001,Sun2005,Cui2007,Tanasittikosol2011}.
Early work on EIA focussed on Zeeman-degenerate systems, where the interpretations of the phenomenon relied upon transfer of coherence \cite{Lezama1999,Taichenachev2000,Goren2004a} or population \cite{Goren2004}, with a minimum of four levels.
More recently, EIA has been observed in a degenerate lambda system \cite{Chanu2012} and also in a four-level $\mathcal{N}$-system \cite{Bason2009}, even when the degeneracy is lifted by applying a small magnetic field.
The effects of thermal motion have also been investigated \cite{Tilchin2011}.

Here we present the first experimental observation of EIA in a non-degenerate three-level ladder system.
In a thermal $^{87}$Rb vapour, the degeneracy is lifted by a strong magnetic field, in which the atoms enter the hyperfine Paschen-Back regime \cite{Sargsyan2012,Zentile2014,Sargsyan2015} where all transitions are separated in frequency by more than the Doppler width.
EIA is observed by detecting the transmission of a weak probe beam~\cite{Sherlock2009} through the atomic vapor which is dressed by both co- and counter-propagating control beams. 
The control beams have the same optical frequency and are resonant with an excited-state transition forming a ladder system similar to standard EIT configurations~\cite{Gea-Banacloche1995}.
A model based on \cite{Kyrola1981} reproduces the experimental results and highlights the prominent role of fast atoms.

\begin{figure}
\includegraphics{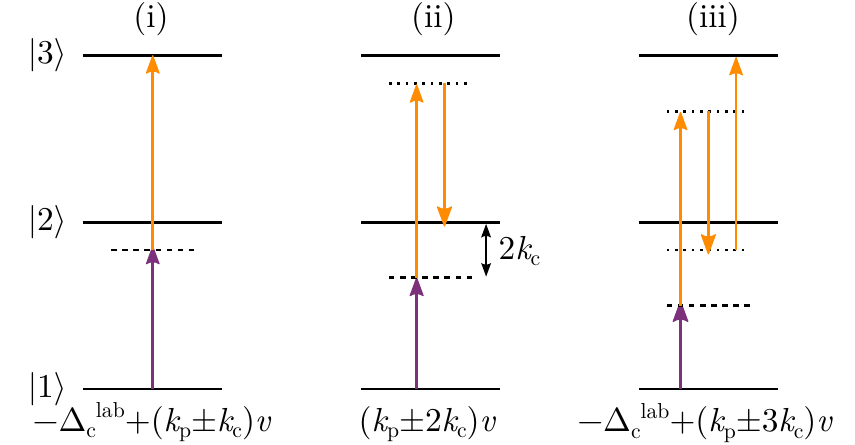}
\caption{\label{multiphoton} (Colour online) Multi-photon resonances. Atoms travelling with velocity component $v$ along the probe field direction observe a control field detuning $\Delta_{\mathrm{c}}^{\mathrm{lab}}\pm k_{\mathrm{c}}v$ for the co- and counter-propagating control fields respectively. 
Odd numbers of control photons (orange) allow resonances between states $\ket{1}$ and $\ket{3}$ (i,iii) and even numbers between states $\ket{1}$ and $\ket{2}$ (ii). 
These resonances occur when the probe photons (purple) are detuned by the amounts shown.}
\end{figure}

The richness of this system comes from multi-photon resonances involving both control beams, that can occur for moving atoms. 
In the rest frame of an atom moving with velocity $v$, the two control fields of wavenumber $k_{\mathrm{c}}$ (with detuning $\Delta_{\mathrm{c}}^{\mathrm{lab}}$ in the laboratory frame) are detuned by $\Delta_{\mathrm{c},\pm}^{\mathrm{atom}} = \Delta_{\mathrm{c}}^{\mathrm{lab}} \pm k_{\mathrm{c}}v$ and thus appear to have a frequency difference $\delta\Delta_{\mathrm{c}}^{\mathrm{atom}}=2k_{\mathrm{c}}v$.
It is therefore possible to form resonances with increasing numbers of control photons, $m$, for a suitable choice of probe detuning $\Delta_{\mathrm{p}}^{\mathrm{lab}}$.
These are shown diagramatically in figure \ref{multiphoton} in the atomic frame.
For odd numbers of control photons (i and iii) we can form resonances between states $\ket{1}$ and $\ket{3}$ for 
\begin{equation}\label{eqn:oddres}
\Delta_{\mathrm{p}}^{\mathrm{lab}}=-\Delta_{\mathrm{c}}^{\mathrm{lab}}+(k_{\mathrm{p}}\pm mk_{\mathrm{c}})v.
\end{equation}
For even numbers of control photons (ii) we can form resonances between states $\ket{1}$ and $\ket{2}$ for 
\begin{equation}\label{eqn:evenres}
\Delta_{\mathrm{p}}^{\mathrm{lab}}=(k_{\mathrm{p}}\pm mk_{\mathrm{c}})v.
\end{equation}
The interaction between these resonances and the main two-level atomic absorption line causes the overall response to differ dramatically from standard EIT.

We adapt the results of \cite{Kyrola1981} to model the probe transmission spectra for the case of a three-level ladder system. 
We consider the situation where the probe and control fields propagate axially along $z$ and drive transitions with Rabi frequencies $\Omega_{\mathrm{p}}(z) = \Omega_{\mathrm{p}}e^{\mathrm{i}k_{\mathrm{p}}z}$ and $\Omega_{\mathrm{c}}(z) = \Omega_{+}e^{-\mathrm{i}k_{\mathrm{c}}z} + \Omega_{-}e^{\mathrm{i}k_{\mathrm{c}}z}$ respectively, where $\Omega_{\pm}$ are the Rabi frequencies associated with the forward- and backward-propagating control beams.
The three-level Bloch equations are solved for the whole medium for atoms in a given velocity class by writing the elements of the density matrix, $\rho$, as a Fourier series expansion in space.
The coherence on the probe transition is then given by $\rho_{12} = e^{\mathrm{i}k_{\mathrm{p}}z}\sum_n\rho_{12}^{(n)}e^{\mathrm{i} n k_{\mathrm{c}}z}$.
Since we are only interested in terms oscillating at the probe frequency, the relevant action of the medium on the probe field is entirely contained within the term $\rho_{12}^{(0)}$. 
This is equivalent to the rotating wave approximation.
The solution for $\rho_{12}^{(0)}$ for a given velocity class $v$ can be written as a continued fraction involving the two Lorentzian lineshapes ($j=2,3$)
\begin{equation}
L_{1j}(m) = \left[\gamma_{1j}+\mathrm{i}\Delta_{j}^{\mathrm{lab}} + \mathrm{i}(k_{\mathrm{p}}+mk_{\mathrm{c}})v\right]^{-1},
\end{equation}
which describe precisely the aforementioned multi-photon resonances, for $m$ control photons.
Here, $\Delta_{2}^{\mathrm{lab}}=\Delta_{\mathrm{p}}^{\mathrm{lab}}$, $\Delta_{3}^{\mathrm{lab}}=\Delta_{\mathrm{p}}^{\mathrm{lab}}+\Delta_{\mathrm{c}}^{\mathrm{lab}}$, $\gamma_{12}$ and $\gamma_{13}$ are the decay rates of the coherences between the corresponding states.

In these terms, the coherence is
\begin{eqnarray}\label{eqn:rho12}
\frac{\rho_{12}^{(0)}}{\Omega_{\mathrm{p}}} &=\mathrm{i}[L_{12}(0)^{-1} +\Omega_{+}^{2}L_{13}(-1) + \Omega_{-}^{2}L_{13}(1) \nonumber\\
&\hspace{.5pt}-X_{+}(1)-X_{-}(1)]^{-1},
\end{eqnarray}
\newline 
where the $X_{\pm}$ functions are given by the recurrence relation
\begin{eqnarray}\label{eqn:rho12X}
X_{\pm}(m) &=& \Omega_{+}^{2}\Omega_{-}^{2}L_{13}(\pm m)^{2} \times \nonumber\\
&&\hspace{.5pt}[L_{12}(\pm m \pm 1)^{-1} + \Omega_{\pm}^{2}L_{13}(\pm m) \nonumber\\
&&\hspace{.5pt} + \Omega_{\mp}^{2}L_{13}(\pm m \pm 2) - X_{\pm}(m+2)]^{-1}.
\end{eqnarray}
The first term of equation (\ref{eqn:rho12}) is the two-level response, while the second and third terms account for the independent effects of the counter- and co-propagating control beams in the three-level system respectively (corresponding to Doppler-free and Doppler-sensitive EIT). 
The additional terms originate from interferences between the two, and their recurrent definition leads to a continued fraction in the full expression. 
Higher $m$ terms in this continued fraction correspond to higher-order interaction of the medium with control photons, and involve higher-order multi-photon resonances ocurring for particular values of $v$ and $\Delta_{\mathrm{p}}^{\mathrm{lab}}$.

The transmission of the probe beam through a thermal medium of length $l$ is given by the Beer-Lambert law with absorption coefficient $\alpha = (-2Nd_{12}/\epsilon_{0})\mathrm{Im}\langle\rho_{12}^{(0)}\rangle_{v}k_{\mathrm{p}}l/\Omega_{\mathrm{p}}$ where $N$ is the atomic number density, $d_{12}$ is the dipole matrix element of the probe transition, $\epsilon_{0}$ is the vacuum permittivity, and $\langle\rho_{12}^{(0)}\rangle_{v}$ is the atomic coherence averaged over the Maxwell-Boltzmann velocity distribution.

\begin{figure}
\includegraphics{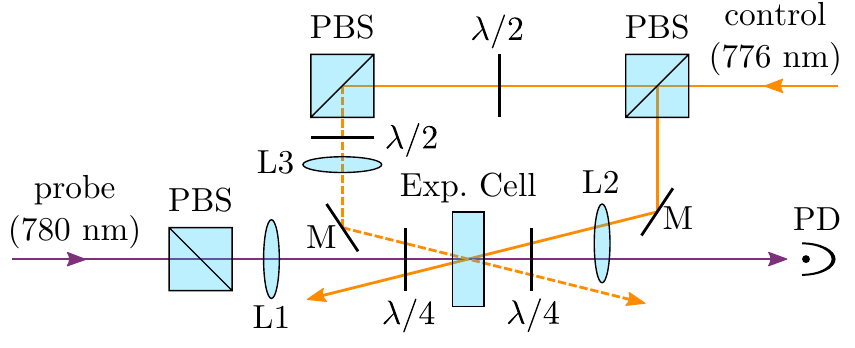}
\caption{\label{setup} Schematic of the experimental setup. PBS - polarising beam splitter; $\lambda/2$ - half-wave plate; $\lambda/4$ - quarter-wave plate; L - lens; M - plane mirror; PD - photodiode. Switching from EIT to EIA is achieved by switching on the dashed 776nm control beam. Angles are not to scale.}
\end{figure}

Experimentally, we have studied the probe field transmission spectra using the setup in figure \ref{setup}.
A weak probe beam (purple) was focussed through a 2 mm long vapour-cell containing $98.2\%$ $^{87}$Rb and $1.8\%$ $^{85}$Rb heated to 80 $^\circ \mathrm{C}$, in a uniform magnetic field of strength $B=0.62$ T. 
In the hyperfine Paschen-Back regime, a single non-degenerate ladder 5S$_{1/2}(m_{J}=1/2,m_{I}=1/2)\rightarrow\mathrm{5P}_{3/2}(m_{J}=3/2,m_{I}=1/2)\rightarrow\mathrm{5D}_{5/2}(m_{J}=5/2,m_{I}=1/2)$ was addressed by using circularly polarised probe (780 nm) and control (776 nm) beams.
We measured the probe transmission as a function of probe detuning in the presence of control light with fixed detuning.
The beam waists ($1/\mathrm{e}^2$) at the centre of the vapour cell were $50$ $\mu \mathrm{m}$ (probe) and $120$ $\mu \mathrm{m}$ (control).

\begin{figure*}
\includegraphics[width=\columnwidth]{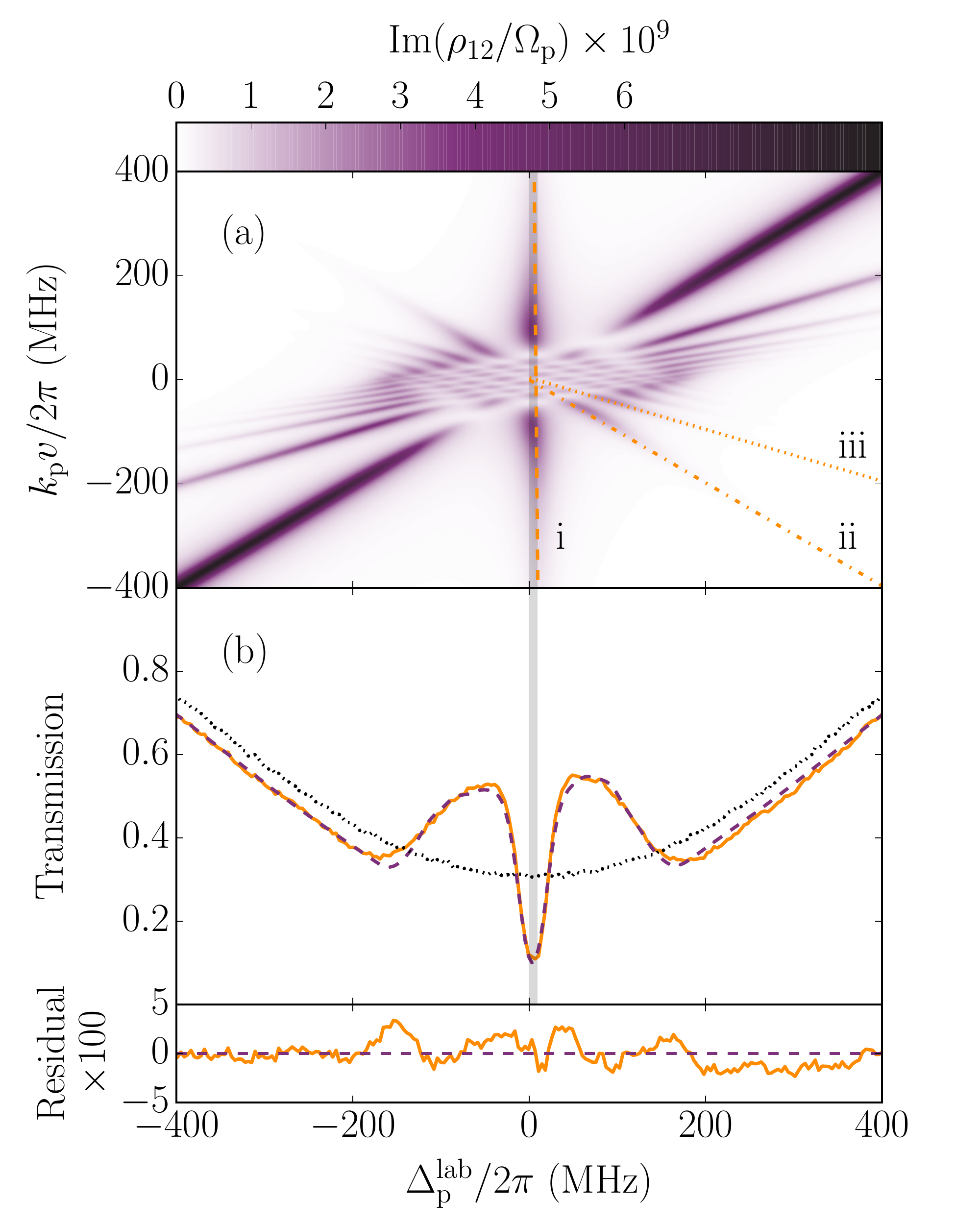}
\includegraphics[width=\columnwidth]{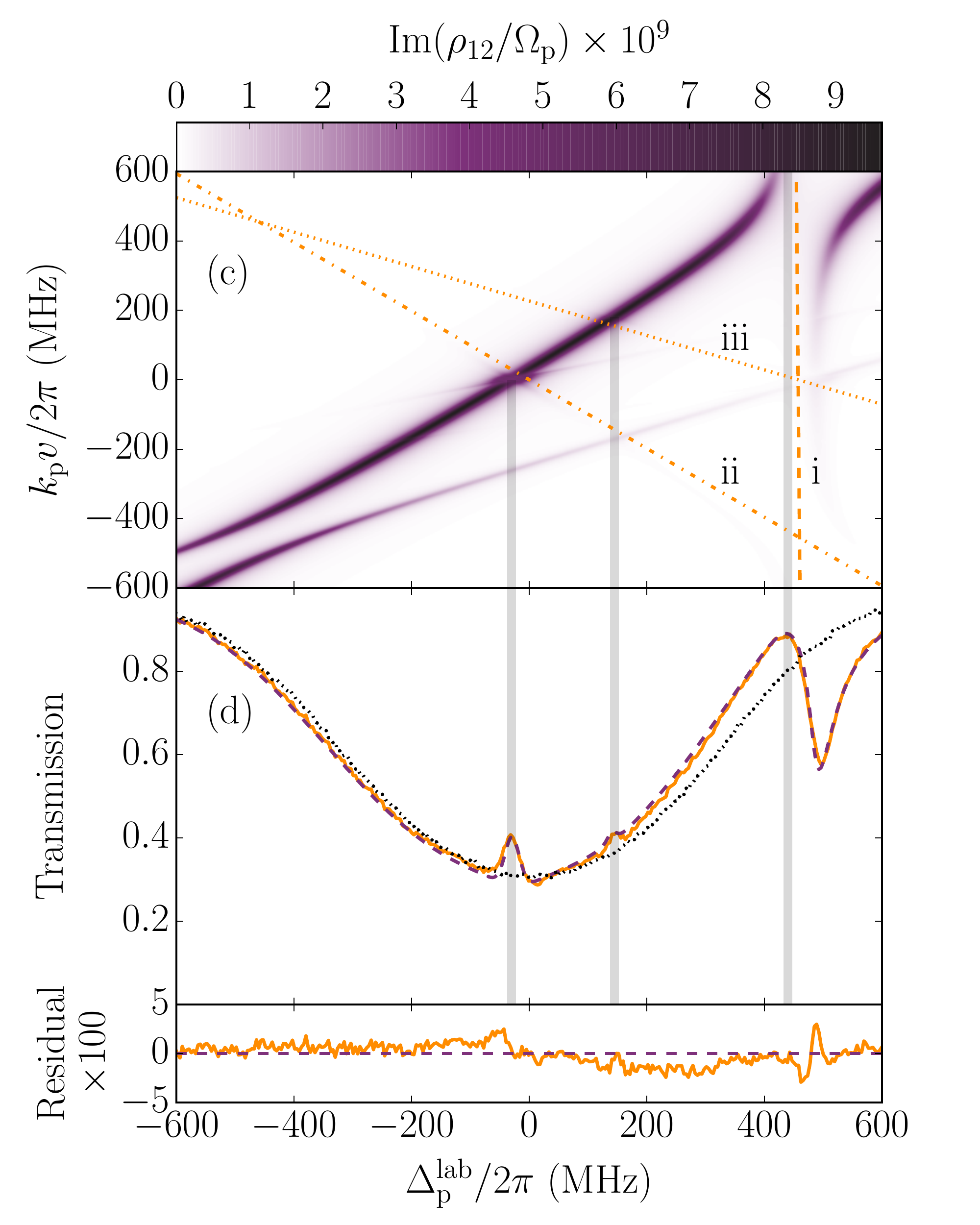}
\caption{\label{sweit} Panels (a) and (c) show the effect on the absorption of different atomic velocity classes produced by a resonant and off-resonant standing wave control field respectively. 
Resonances associated with 1, 2 and 3 control photons are shown for positive detunings by the lines labelled (i), (ii) and (iii) respectively. 
The crossing of these resonances with each other and the main two-level absoption line produces features such as enhanced absorption in (a) and multiple transparencies in (b) at the detunings indicated by the vertical grey lines.
Panels (b) and (d) show the probe transmission spectra which are obtained by integrating panels (a) and (c) over all velocities weighted by the Maxwell-Boltzmann distribution. 
The dashed (purple) line shows the result of numerically fitting the theoretical model (equation (\ref{eqn:rho12})) to the experimental data (solid orange line) with the residuals, plotted below, showing excellent agreement. 
The parameters chosen for the colour-maps in (a) and (c) are the result of this numerical fit, where all parameters are constrained to be the same for the two spectra except for the control detuning. 
The resulting parameters are $\Omega_+/2\pi = 87$ MHz, $\Omega_-/2\pi = 83$ MHz, $\gamma_{12}/2\pi = 15$ MHz, $\gamma_{13}/2\pi = 16$ MHz, $\Delta_{\mathrm{c}}^{\mathrm{lab}}/2\pi = -8$ MHz in (b) and $\Delta_{\mathrm{c}}^{\mathrm{lab}}/2\pi = -457$ MHz in (d). 
The dotted (black) line shows the transmission spectrum with both control beams switched off, from which the atomic number density can be extracted using the ElecSus code \cite{Zentile2015}.}
\end{figure*}

First, we consider the case of a resonant standing-wave control field ($\Delta_{\mathrm{c}}^{\mathrm{lab}}=0$).
Figure \ref{sweit}(a,b) shows its effect on the absorptive response of different velocity classes of atoms (calculated from equation (\ref{eqn:rho12})), and the resulting probe transmission spectrum (experimental and theoretical). 
Most noticeably, the Doppler-free EIT window opened by the counter-propagating control beam is modified by the addition of the second control beam, producing a narrow and strong enhanced absorption feature on resonance (solid orange line compared to dotted black line in panel (b)).
We use a Marquardt-Levenberg fitting routine \cite{Hughes2010} to fit the theoretical model (equation (\ref{eqn:rho12})) to the data and find excellent agreement, with typical RMS residuals of $<0.01$.
We choose to truncate the continued fraction at $m=7$ as further increasing the number of terms included does not produce a noticeable change in the resulting lineshape.
The velocity map (figure \ref{sweit}(a)) exhibits the superposition of the strong one-photon absorption line ($\Delta_{\mathrm{p}}^{\mathrm{lab}} = k_{\mathrm{p}}v$) with all previously discussed resonances (see equations (\ref{eqn:oddres},\ref{eqn:evenres})), crossing at $\Delta_{\mathrm{p}}^{\mathrm{lab}}=0$. 
In particular, it shows that the resonant EIA is due to an integrated effect of all velocity classes having strong absorption at $\Delta_{\mathrm{p}}^{\mathrm{lab}}=0$. 
A complementary physical explanation for this is as follows.
For a fast atom such that $k_{\mathrm{c}} v\gg\Omega_{\pm}$, the two resonant control beams appear to have large detunings  $\Delta_{\mathrm{c},\pm}^{\mathrm{atom}} = \pm k_{\mathrm{c}}v$. 
Taken independently they would thus give rise to AC Stark shifted two-photon absorption lines at probe detunings of $\Delta_{\mathrm{p}}^{\mathrm{lab}} \approx \mp \Omega_{\pm}^{2}/4k_{\mathrm{c}}v$ \cite{Fleischhauer2005}.
Therefore to a first order approximation the addition of the second control beam simply cancels the light-shift of the first, creating a 2-photon resonance at precisely $\Delta_{\mathrm{p}}^{\mathrm{lab}}=0$ for all velocity classes.
In figure \ref{sweit} the control field was detuned by a small amount ($\Delta_{\mathrm{c}}^{\mathrm{lab}}/2\pi= -8$ MHz) so that the center of the enhanced absorption is also slightly detuned.

We now consider the effect of an off-resonant control field that is detuned by $\Delta_{\mathrm{c}}^{\mathrm{lab}}/2\pi=-457$ MHz.
The velocity map and associated transmission spectrum are shown in figure \ref{sweit}(c,d).
In this case, the most visible features on the velocity map are the two large avoided crossings on the one-photon probe absorption line, around  $\Delta_{\mathrm{p}}^{\mathrm{lab}} \approx \pm \Delta_{\mathrm{c}}^{\mathrm{lab}}$. 
These correspond to the EIT effects of the two control beams considered independently (Doppler-free or not).
In addition several distinct transparencies appear (shown by the vertical grey lines) due to interference between the one- and many-photon resonances. 
For an off-resonant control field the light-shifts due to counter-propagating beams do not cancel.
Instead, where the multi-photon resonances coincide with the one-photon absorption line, there are avoided crossings leading to small windows of transparency.
In terms of these multi-photon transitions, the simple two-photon resonance (i), which is the usual feature associated with off-resonant Doppler-free EIT, corresponds to the crossing with $\Delta_{\mathrm{p}}^{\mathrm{lab}}=-\Delta_{\mathrm{c}}^{\mathrm{lab}}+(k_{\mathrm{p}} - k_{\mathrm{c}})v \approx -\Delta_{\mathrm{c}}^{\mathrm{lab}}$.
The three-photon resonance (ii) occurs when $\Delta_{\mathrm{p}}^{\mathrm{lab}}=(k_{\mathrm{p}} - 2k_{\mathrm{c}})v$. 
More generally, resonances involving even number of control photons do not depend on the control detuning, and all cross at $\Delta_{\mathrm{p}}^{\mathrm{lab}}=0$ (see equation (\ref{eqn:evenres})).
As a result, we predict and observe a probe transparency around resonance, for arbitrary control detunings.
Experimentally, we see that the transparency does not occur exactly at $\Delta_{\mathrm{p}}^{\mathrm{lab}}=0$.
This can be understood as the light-shift of state $\ket{2}$ due to coupling of the states $\ket{2}$ and $\ket{3}$ by the off-resonant control field.
We therefore expect the shift to reduce with increasing control detuning, which is observed experimentally.
The small resonance (iii) is associated with a four-photon transition occuring when $\Delta_{\mathrm{p}}^{\mathrm{lab}}=-\Delta_{\mathrm{c}}^{\mathrm{lab}}+(k_{\mathrm{p}} - 3k_{\mathrm{c}})v$, and although there are in theory infinitely many resonances involving more than three control photons, in practice their amplitude very quickly becomes negligible. 

In conclusion we have observed EIA for the first time in a non-degenerate three-level ladder system, demonstrating that a 4-level system is not required to observe EIA.
Our model, adapted from \cite{Kyrola1981}, shows excellent agreement with experimental transmission spectra and clearly exposes the origin of enhanced absorption in these systems where fast velocity classes contribute strongly to the overall absorption.
We therefore predict that resonant EIA will not be observed for three-level cold atoms but even so that the EIT spectra will be significantly modified by the presence of a second control beam. 

We acknowledge financial support from EPSRC (grant EP/L023024/1) and Durham University.
The data presented in this paper are available at \url{http://dx.doi.org/10.15128/kk91fk98z}.

\bibliography{references}

\newpage

\end{document}